\documentclass[sigconf]{acmart}

\AtBeginDocument{%
  \providecommand\BibTeX{{%
    \normalfont B\kern-0.5em{\scshape i\kern-0.25em b}\kern-0.8em\TeX}}}

\usepackage[utf8]{inputenc}

\usepackage{balance}
\usepackage{graphicx,subfigure}
\usepackage{float}
\usepackage{epstopdf}
\usepackage{datetime2}
\usepackage{bbding}
\usepackage{array}
\usepackage{listings}
\usepackage{soul}
\usepackage{color}
\usepackage{listings}
\usepackage{courier}
\usepackage{tcolorbox}
\usepackage{multirow}
\usepackage{booktabs}
\usepackage{makecell}
\usepackage{rotating}
\usepackage{appendix}

\newcommand{\http}[1]{{\color{red}#1}}

\usepackage{url}

\newcolumntype{P}[2]{%
  >{\begin{turn}{#1}\begin{minipage}{#2}\small\raggedright\hspace{0pt}}l%
  <{\end{minipage}\end{turn}}%
}

\makeatletter
\g@addto@macro{\UrlBreaks}{\UrlOrds}
\makeatother

\lstdefinelanguage{MXML}{
    alsoletter={-,:},
    basicstyle=\scriptsize\tt,
    numberstyle=\scriptsize\color{red} \tt,
    keywords={},
    keywordstyle=[1]\color{teal}\bfseries,
    keywords=[2]{POST, SOAPAction:},
    keywordstyle=[2]\color{darkgray}\bfseries
}
\lstdefinelanguage{pseudo}{
    basicstyle=\scriptsize\tt,
    numberstyle=\scriptsize\color{red} \tt,
    keywords={func},
    keywordstyle=[1]\color{darkgray}\bfseries,
    keywords=[2]{onStart, MD5, extract_from_preferences, read_file, parse_tickets, delete_history, HMACSHA1, encrypt, AES, base64_encode},
    keywordstyle=[2]\color{teal}\bfseries,
    keywords=[3]{function, if, else, return},
    keywordstyle=[3]\color{brown}\bfseries,
    commentstyle=\itshape\color{gray},
    comment=[l]{\#}
}
\lstdefinelanguage{Java}{
    alsoletter={-},
    basicstyle=\scriptsize\tt,
    numberstyle=\scriptsize\color{red} \tt,
    keywords=[1]{int, double, DateTime, string, internal, static, byte},
    keywordstyle=[1]\color{blue}\bfseries,
    keywords=[2]{for, if, else, try, catch, this, return, public},
    keywordstyle=[2]\color{teal}\bfseries
}
\lstdefinelanguage{NONE}{}
\begin{document}

\title{Android OS Privacy Under the Loupe -- A Tale from the East}

\author{Haoyu Liu}
\affiliation{%
  \institution{The University of edinburgh}
  \city{Edinburgh}
  \country{United Kindgom}}
\email{haoyu.liu@ed.ac.uk}

\author{Douglas J. Leith}
\affiliation{%
  \institution{Trinity College Dublin}
  \city{Dublin}
  \country{Ireland}}
\email{Doug.Leith@tcd.ie}

\author{Paul Patras}
\affiliation{%
  \institution{The University of edinburgh}
  \city{Edinburgh}
  \country{United Kindgom}}
\email{paul.patras@ed.ac.uk}

\renewcommand{\shortauthors}{Haoyu Liu, Douglas J. Leith, \& Paul Patras}

\begin{abstract}
China is currently the country with the largest number of Android smartphone users.  We use a combination of static and dynamic code analysis techniques to study the data transmitted by the preinstalled system apps on Android smartphones from three of the most popular vendors in China. We find that an alarming number of preinstalled system, vendor and third-party apps are granted dangerous privileges. Through traffic analysis, we find these packages transmit to many third-party domains privacy sensitive information related to the user's device (persistent identifiers), geolocation (GPS coordinates, network-related identifiers), user profile (phone number, app usage) and social relationships (e.g., call history), without consent or even notification. This poses serious deanonymization and tracking risks that extend outside China when the user leaves the country, and calls for a more rigorous enforcement of the recently adopted data privacy legislation.
\end{abstract}



\maketitle
\section{Introduction}
As of 2021, China is the country with the largest number of smartphone users~\cite{newzoo}, with over 70\% of mobile handsets running an Android OS distribution~\cite{statcounter}. Concerns about user and data privacy on mobile handsets are increasingly coming to the attention of regulatory authorities in many countries, including in the EU~\cite{gdpr}, Canada~\cite{pipeda}, USA~\cite{nyt}, Brazil~\cite{lgpd}, and Japan~\cite{appi}. While China has recently adopted a Personal Information Protection Law \cite{pip-cn},  which mirrors in part the EU GDPR, the extent to which smartphone vendors comply with these provisions and how much sensitive personal information may be disclosed to vendors and third parties without user consent remains poorly understood. Prior privacy analysis of preinstalled Android system apps~\cite{our-report} has largely been confined to European handsets and does not consider Android OS distributions used by consumers in China, where regional differences are directly reflected in the firmware and built-in system apps.

In this paper, we conduct an in-depth  analysis of the Android OS variants from three of the most popular smartphones vendors in China, namely OnePlus, Xiaomi and Oppo Realme. We analyze the OS system apps and their permissions, as well as the communication between {these system apps} and the servers to which these connect in order to provide users with the intended functionality. We confine consideration to the information transmitted by the OS and system apps that come preinstalled on a handset, leaving out any third-party apps that a user would install themselves. We note that system apps such as Settings, Messages, and Maps cannot be removed by the user, and preinstalled system apps can have access privileges that user-installed apps may not gain easily, especially not without explicit user consent.

We focus on a privacy-aware but busy user who opts out of analytics and personalization, does not use any cloud storage or any other optional third-party services, and has not set up an account on any platform of the OS distribution developer. This allows us to establish a baseline for privacy behavior.  We find that the smartphones studied send a worrying amount of Personally Identifiable Information (PII) not only to the device vendor but also to Chinese mobile network operators (e.g., China Mobile and China Unicom), even though they do not provide any service to the device, i.e., a SIM card has not been inserted or we use a SIM card that ensures connectivity to a different operator in China or in the UK, and to over-the-top service providers (e.g., Baidu). The data we observe being transmitted includes persistent device identifiers (IMEI, MAC address, etc.), location identifiers (GPS coordinates, mobile network cell ID, etc.), 
user profiles (phone number, app usage patterns, app telemetry), and social connections (call/SMS history/time, contact phone numbers, etc.). Combined, this information poses serious risks of user deanonymization and extensive tracking, particularly since in China every phone number is registered under a citizen ID. Moreover, the data collection behaviors do not change when the devices move outside China, despite potentially being under jurisdictions where users should benefit from stronger data protection, meaning that phone vendors and some third-parties are still able to track business travelers and students studying abroad, including the foreign contacts they make on their visits.

Finally, we perform a cross-regional analysis and compare the preinstalled system apps on the Chinese (CN) and Global (e.g., EU) Android OS distributions from the same OS developers. We find that the number of preinstalled third-party apps on CN OS distributions is 3 to 4 times larger than for the corresponding Global OS distribution, and that these are given 8 to 10 times as many permissions as third-party apps in Global distributions, including many more permissions classed as dangerous. Overall, our findings paint a troubling picture of the state of user data privacy in the world's largest Android market, and highlight the urgent need for tighter privacy controls to increase the ordinary people's trust in technology companies, many of which are partially state-owned.










\vspace{-1em}
\section{Threat Model}
We focus on privacy leaks from system services and preinstalled system apps that transmit information to backend servers. Unlike third-party apps which have to explicitly ask for system permissions (storage, location, network, etc.), preinstalled system apps may use permissions, including privileged permissions not available to non-system apps, without asking for user consent. 

\begin{figure*}[t]
    \centering
    \includegraphics[width=0.85\linewidth]{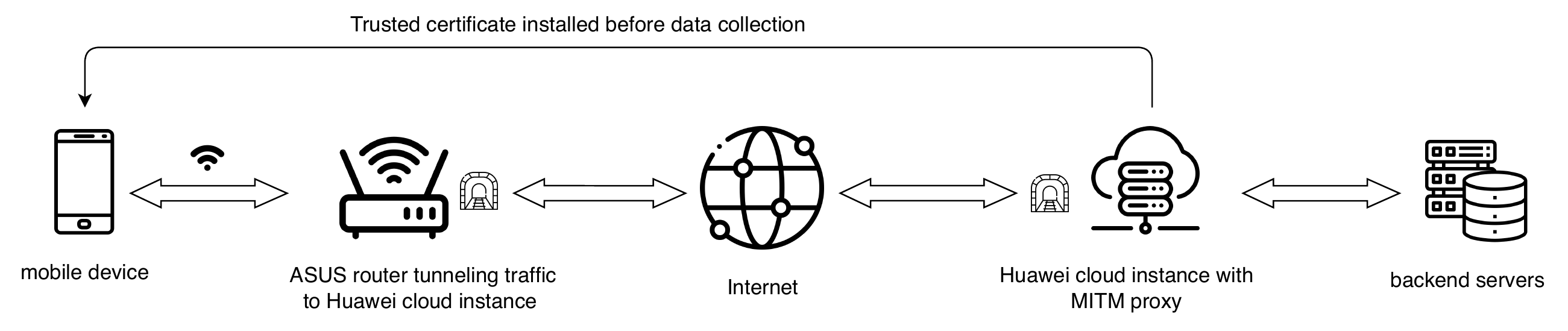}
     \vspace{-1.5em}
    \caption{Illustration of the experimentation setup. A wireless router is configured to tunnel all traffic from the connected mobile devices to a Huawei Cloud instance in Shanghai, where a \textit{mitmproxy} is set up to intercept and log HTTPS flows. A \textit{MTIMProxy} signed certificate is installed on every handset studied prior to any data collection.}
    \label{fig:setup}
    \vspace{-1.5em}
\end{figure*}

It is worth noting that data transmission from the OS does not intrinsically entail a breach of privacy. For instance, it can be useful to share details of the device model/version and the locale/country of the device when checking for software updates. This poses few privacy risks if the data is common to many handsets and therefore cannot be easily linked back to a specific handset/person. 

Two major issues in handset privacy are (i) release of sensitive data, and (ii) handset deanonymization, i.e., linking of the handset to a person's real world identity.  

\textit{Release of sensitive data}. What counts as sensitive data is a moving target, but it is becoming increasingly clear that data can be used in surprising ways and that so-called metadata can be sensitive data. The sensitive information we consider in this work can be classified into four major categories:
\begin{enumerate}
    \item \textbf{Device-specific:} Information that is bundled with the device upon manufacturing or setup, such as International Mobile Equipment Identity (IMEI), Mobile Equipment Identifier (MEID), Android ID, MAC address, hardware serial number, installed packages and hardware model.
    \item \textbf{Location-specific:} Information that directly or indirectly reveals the location of the user, such as GPS coordinates, the SSID and MAC address of nearby Wi-Fi access points, Mobile Country Code (MCC), Mobile Network Code (MNC), Location Area Code (LAC), and cell ID (CID).
    \item \textbf{User Profile:} Information that reveals user identity (phone number) and user traits (list of installed apps, app usage~logs).
    \item \textbf{Social Relationships:} Information that contains details about the personal contacts of the user, such as phone call and SMS history.
\end{enumerate}
We also note that data which is not sensitive in isolation can become sensitive when combined with other data, see for example~\cite{cominelli2020even,di2017consensus,yen2012host}. This is not a hypothetical concern since vendors such as Xiaomi operate mobile payment services and supply custom web browsers with the handsets they commercialize.  

\textit{Deanonymization}. Android handsets can be directly tied to a person's real identity in at least three ways, even when a user takes active steps to try to preserve their privacy.  Firstly, via the SIM and phone number.   In China every phone number is registered under a citizen ID.    Secondly, via the IMEI number which globally uniquely identifies each SIM slot on a handset.    The IMEI is used by cellular operators to block network access for a stolen handset\ \cite{imei_usage} and so it is commonly linked to the user's SIM, phone number and cellular contract.   Thirdly, via vendor accounts.  For example, Xiaomi encourages people to create user accounts and sign in to them while using their handsets.   Creation of user accounts typically requires disclosure of personal information that is then linked to device identifiers such as the IMEI, and if used for mobile payments becomes linked to a person's financial details (e.g., credit card).  


\section{Ethical Statement}
Our study involved no human subjects and took measurements on handsets in normal operation.  The methodology used therefore raises no ethical concerns. However, the software that we study is in active use by many millions of people, and so warrants consideration of responsible disclosure. We have informed Oppo, Realme and Xiaomi of our findings. It seems likely that any changes to the OSes studied here, even if they were agreed upon, would take a considerable time to deploy and we believe that keeping handset users in the dark for a long open-ended period in the meantime is inappropriate.

\vspace{-0.5em}
\section{Environment Setup}
We scrutinize three popular mobile devices purchased in China, which run  local (CN) firmware, namely a Xiaomi \textit{Redmi Note 11} with Android 11/MIUI 12.5.4.0 RGBCNXM, an OPPO \textit{Realme Q3 Pro} with Android 11/realme UI v2.0 RMX2205\_11\_A.13 (based on ColorOS 11), and a \textit{OnePlus 9R} with Android 11/ColorOS 11.2 LE2100\_11\_A.05. 

Regional differences take the form of differences in the installed firmware, e.g., differences between the system apps installed in the Global/EU vs. CN Android OS distributions.  In addition, even when the same version of system app is installed in the CN and Global distributions, the app may contain logic that causes it to behave differently depending on the region, e.g., by checking the current location. Regarding the latter, by reverse-engineering the main system apps, we find that mobile devices tend to use locale, firmware version, IP address and MCC (which identifies the mobile operator country) to localize their behavior.  For example, Xiaomi maintains a list of URLs in a Java HashMap:
\begin{lstlisting}
 {"CN": "data.mistat.xiaomi.com",
 "INTL": "data.mistat.intl.xiaomi.com",
 "IN": "data.mistat.india.xiaomi.com"},
\end{lstlisting}
and uses a hardcoded value (\texttt{"CN"}) and locale together, to select an endpoint in the Chinese firmware. The preinstalled Amap package on Realme and OnePlus handsets uses the MCC to determine the API endpoint:
\begin{lstlisting}[language=Java]
StringBuilder sb = new StringBuilder();
if (a2) {
    sb.append("http://aps.oversea.amap.com/APS/r?ver=5.1&q=0");
} else if (z) {
    sb.append("http://aps.testing.amap.com/APS/r?ver=5.1&q=0");
} else if (z2) {
    sb.append("https://aps.amap.com/APS/r?ver=5.1&q=0");
} else {
    sb.append("http://aps.amap.com/APS/r?ver=5.1&q=0");}
\end{lstlisting}
in which \texttt{a2} is \texttt{True} if MCC does not equal 460 (China). 

Although the mobile devices in our study were purchased on the Chinese market, we do not conduct experiments locally and recognize that the data collected outside China may not fully represent a device's behavior in that region. To overcome this issue, we set up a network tunnel between our campus and a Huawei Cloud instance in Shanghai. 
The IP address observed by the backend server is thus that of the Huawei Cloud server located in Shanghai. We set up each handset using Chinese as the language to simulate a local user. With this setup, the only app that we observed still adapting to the device's true location was the Amap app. Fortunately, through reverse-engineering (decrypting/decoding every connection in the collected traces via mitmproxy and apk decompiling and analysis, which allows us to examine all the transmitted fields in plaintext) we confirmed that the contents of the messages transmitted to Amap's different API endpoints are the same in all regions, and so leaves our traffic collection unaffected. We did not find any system services or preinstalled applications that use GPS to select an endpoint.  

\begin{table*}[t]
\bgroup
\def\arraystretch{1.1}
\begin{tabular}{@{}ccccccccccc@{}}
\Xhline{2\arrayrulewidth}
\multirow{2}{*}{Region} & \multirow{2}{*}{Model} & \multicolumn{3}{c}{android pkg} & \multicolumn{3}{c}{vendor pkg} & \multicolumn{3}{c}{3rd-party pkg} \\ \cmidrule(l){3-5} \cmidrule(l){6-8}  \cmidrule(l){9-11} 
 &  & num & \# perm & \begin{tabular}[c]{@{}c@{}}\# dangerous \\ permission\end{tabular} & num & \# perm & \begin{tabular}[c]{@{}c@{}}\# dangerous \\ perm\end{tabular} & num & \#perm & \begin{tabular}[c]{@{}c@{}}\# dangerous \\ perm\end{tabular} \\ \midrule
\multirow{3}{*}{CN} & Redmi & 152 & 36.0 & 3.1 & 137 & 117.8 & 8.9 & 36 & 72.9 & 7.1 \\
 & OnePlus & 159 & 32.8 & 3.2 & 185 & 72.9 & 6.7 & 40 & 51.9 & 6.5 \\
 & Realme & 147 & 35.6.3 & 3.7 & 163 & 71.5 & 7.8 & 34 & 56.2 & 6.5 \\ \midrule
\multirow{2}{*}{Global} & Redmi & 169 & 32.4 & 3.4 & 111 & 108 & 9.2 & 14 & 9.0 & 1.8 \\
 & Realme & 166 & 35.1 & 4.1 & 139 & 100.0 & 9.9 & 9 & 6.2 & 4.1 \\  \Xhline{2\arrayrulewidth}
\end{tabular}
\egroup
\caption{Summary of the number of different types of packages installed in each of the handsets studied, and the average number of permissions requested in each category. Note that we group hardware-supported packages, including Qualcomm, Mediatek, and packages developed by parent/child companies into the `vendor pkg' category.}
\label{table:pkg_num}
\vspace{-2.5em}
\end{table*}

\vspace*{-1em}
\subsection{Wi-Fi Connection \& Traffic Tunneling}
{We} aim to capture the handset traffic starting from right after factory reset.  To collect measurements, we first configured a Wi-Fi access point on an ASUS RT-AC86U router which supports third-party firmware and thus allows for the configuration of many VPN or tunneling protocols.
We build a tunnel where the proxy server is based on a Huawei Cloud s6.small instance running Ubuntu 20.04 in Shanghai, and the ASUS router running Koolshare 3.0.0.4 works as a proxy client, which is configured to redirect any TCP/UDP packets to the endpoint on the Huawei Cloud instance (see Figure~\ref{fig:setup}) To avoid any negative impact on the censorship circumvention community, we do not disclose the tunneling protocol used. During our traffic collection campaign, only the handsets being studied are permitted to connect to the Wi-Fi access point, in order to prevent other sources of traffic from interfering with our measurements.

\vspace*{-1em}
\subsection{Man-in-the-Middle Proxy}
The tunneling server receives connections from the handset and forwards them to the intended destinations, while we deploy a man-in-the-middle proxy to be able to intercept and decipher HTTP/HTTPS traffic. 
To fully isolate handset-initiated requests from the Huawei Cloud messaging that serves to monitor the hosted virtual machine (VM), we create a user named \texttt{tunnel} which runs the tunneling proxy server. We also run \textit{mitmproxy} 8.0.0 \cite{mitmproxy} with superuser permissions on port 8080 on the VM, and configure \texttt{iptables} to redirect any TCP connections from \texttt{tunnel} to \texttt{locahost:8080}. In this way, \textit{mitmproxy} communicates with the handset on behalf of the requests from the server endpoints, and initiates new requests to the target server endpoints by impersonating the handset, which allows \textit{mitmproxy} to intercept each request. 

Due to the prevalence of HTTPS, the mobile device needs to be equipped with a \textit{mitmproxy}-generated CA certificate, so that TLS handshakes between the handset and \textit{mitmproxy} can succeed. However, if installing the \textit{mitmproxy} certificate as a user-trusted certificate, the user has to authenticate by PIN-code or password every time the certificate is used. Therefore, we install it as a system-trusted certificate instead, following the steps explained in \cite{our-report}, which requires the handsets to be rooted. Xiaomi, OPPO and OnePlus are open about unlocking the bootloader of their devices, which allows us to flash Magisk-patched boot images. 
In our work, we use Magisk v23.0 \cite{magisk} to unpack the stock boot image and hijack the original SELinux policy in order to acquire root access.

\vspace{-1em}
\section{Datasets}
\label{sec:dataset}
\subsection{Experiment Design}
We collect traffic generated by the three mobile devices considered (Xiaomi Redmi Note 11, OPPO Realme Q3 Pro, and OnePlus 9R), which run firmware made for the Chinese market. We root each device and perform factory reset prior to any data collection. During the device setup stage, the user is asked to agree with terms and conditions, and to customize a set of options. Similar to prior work \cite{our-report}, we act as a privacy-aware user and uncheck all the options presented by the OS. For \textit{Redmi Note 11}, we uncheck `Turn On Screen to Activate Mi AI', `Location', `Send Usage and Diagnostic Data', `Automatic System Updates', `User Experience Program', and skip System Navigation Mode; for \textit{Realme Q3 Pro}, we uncheck `Join User Experience Program', `Automatically Select the Best Wi-Fi', `Automatically Switch to Mobile Network', `Lock Screen Magazine', `Auto Update Overnight', and skip Learn Swipe Gestures; for \textit{OnePlus 9R}, we uncheck `Join User Experience Program', `Automatically Select the Best Wi-Fi', `Automatically Switch to Mobile Network', `Auto Update Overnight' and skip Learn Swipe Gestures.

During the setup stage, we also skip the Wi-Fi connection configuration and do not insert a SIM card, to make sure that no Internet traffic is generated. After installing the \textit{mitmproxy} certificate, we perform the following set of actions on each handset:
\vspace{-0.5em}
\begin{enumerate}
    \item Connect to the controlled Wi-Fi access point;
    \item Leave the handset untouched for 24 hours and log the network activity via \textit{mitmproxy} on Huawei Cloud;
    \item Insert a Chinese SIM card, disable Mobile Data and log the network activity through Wi-Fi. The phone was connected to a UK cellular network at this stage;
    \item Keep the handset idle for 24 hours with a SIM card inserted and record the network activity;
    \item Turn off and on location and record the network activity;
    \item Make and receive a phone call, and log the network activity;
    \item Send and receive a message, and log the network activity;
    \item Open Camera, Clock, Note, Photos, Recorder, Settings one-by-one, and log the network activity. If any application asks for permissions, we only grant the minimal set that ensures they remain functional;
    \item Remove the SIM card and log the network activity;
    \item Factory Reset the device and log the network activity.
\end{enumerate}
In addition, to allow a cross-regional comparison we obtained the traces collected by the authors of \cite{our-report} from Xiaomi \textit{Redmi Note 9} with Android 10/MIUI Global 12.0.7 QJOMIXM (denoted as Redmi Global) and OPPO \textit{Realme 6 Pro} with Android 10/realme UI v1.0 RMX2063\_11 \_A.38 (denoted as Realme Global), where both run the global versions of the respective firmware. 
Note that the traces collected from the CN firmware do not contain any connections to Google services, and thus in our study we filter out the Google connections in Redmi Global and Realme Global for a fair comparison.  

\vspace*{-1em}
\subsection{Limitations}
We highlight a number of potential limitations of the traces collected in our study. 

1) We do not alter the GPS location of the handsets and we do not know whether being inside mainland China would trigger the AMap maps app to behave differently. This is limited to the Realme and OnePlus handsets, which have the AMap app preinstalled and running in the background. 

2) Due to the passage of time since the measurements reported in~\cite{our-report}, the devices in our investigation are shipped with Android 11, whereas the devices in \cite{our-report} run Android 10. However, we find (see Section \ref{sec:traffic_analysis}) that the differences that we observe between the CN and Global variants of the Android OS are not due to the Android version, but instead are mainly associated with the preinstalled applications and system services.

3) We only collect and analyze the network traffic generated by system apps and by basic applications such as the dialer and messages apps.  Nevertheless, we argue that these largely reflect the attitude of vendors towards user privacy. 

4) Traffic analysis is limited to the apps which do not implement certificate pinning. 
\vspace*{-1em}
\subsection{Additional Material: Connection Data}

Annotated measurements of network connections are available at \url{https://github.com/Mobile-Intelligence-Lab/android_CN_trafficdata}.

\begin{figure*}[t]
    \centering
    \vspace*{-1em}
    \includegraphics[width=\linewidth]{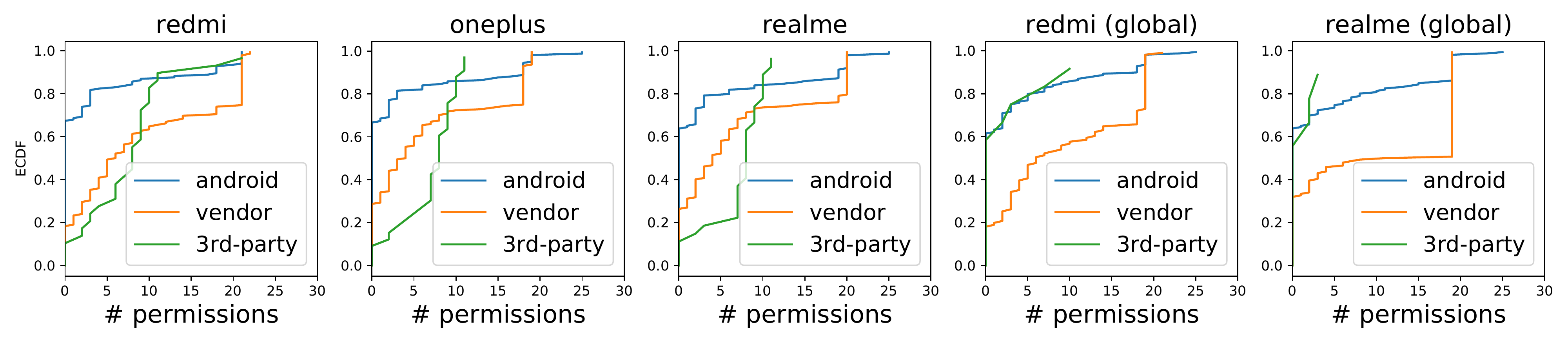}
    \vspace*{-2.5em}
    \caption{The ECDF of the number of dangerous permissions requested by each category of packages in each handset.}
    \label{fig:perm_ecdf}
    \vspace*{-1.5em}
\end{figure*}

\section{Results and Analysis}
We carry out our privacy analysis using a mix of static and dynamic analysis techniques:  1) static analysis is used to study the differences between installed packages and the permissions requested in each handset; and 2) dynamic analysis via code instrumentation and traffic analysis is used to discover whether sensitive user information is actually leaked to backend servers. We seek to answer the following key questions:
\begin{enumerate}
    \item How many apps are preinstalled in each handset? 
    \item How many permissions are requested by system services and preinstalled third-party apps?
    \item Do handset manufacturers grant runtime permissions by default to preinstalled third-party packages?
    \item Which type of personally identifiable information (PII) is uploaded to backend servers?
    \item Is the user notified about the transmission of the PII regarding geolocation, user profile and social relationships?
    \item Who collects PII and where are they located?
\end{enumerate}
The first three questions aim to compare the differences between the Android distributions on the handsets from a static perspective, while the latter three focus on the dynamic aspects. The static analysis is crucial to understanding how preinstalled apps are able to achieve exfiltration, especially of some PII that is supposed to be protected under runtime permissions. Our traffic analysis reveals the transmissions of a collection of PII while users may be completely unaware of this.

\vspace{-0.5em}
\subsection{Static Analysis}
Third-party app developers often collaborate with mobile device manufacturers, who embed popular third-party applications in the official firmware for a specific region. For example, it is common for Chinese brands to preinstall Chinese input apps, video streaming apps (such as Youku and Tencent TV), and domestic map apps (e.g., Baidu Map and AMap), due to the governmental ban on Google services. From a user perspective, this constitutes product bundling and preinstalled applications may excessively request permissions without the user knowing it. 

\vspace{-0.5em}
\begin{tcolorbox}
Q1. How many apps are preinstalled in each handset?
\end{tcolorbox}
The entire list of installed Android packages is maintained in \texttt{/data/system/packages.list}, and the requested permissions and whether each permission is granted, can be acquired by running \texttt{dumsys package <packgename>} on Android devices. Table \ref{table:pkg_num} shows the details of the preinstalled packages found on the devices we study, and the average number of requested (dangerous) permissions encountered on each. Specifically, we group preinstalled packages into three categories, Android AOSP packages, vendor packages and third-party packages. It is the vendor who decides which type of hardware (CPU, fingerprint sensor, etc.) to use during production, thus we count hardware-supported packages including Qualcomm and Mediatek in the `vendor pkg' category. Moreover, we also group the packages developed by parent/child companies into the `vendor pkg' class. We find that the number of preinstalled packages among different brands with Chinese firmware is roughly the same, with vendor packages on OnePlus slightly outnumbering that on the other two brands. This is because OnePlus shares the same Android OS distribution (i.e., ColorOS) with Realme, but also loads a number of OnePlus self-developed apps. There are more than 30 third-party packages deployed in each handset with Chinese firmware, including multiple similar types of application. For example, \textit{Redmi Note 11} is bundled with three Chinese input apps, namely, Baidu Input, IflyTek Input and Sogou Input. Both \textit{OnePlus 9R} and \textit{Realme Q3 Pro} preinstall Baidu Map as a foreground navigation app but also load the AMap package, which is continuously running in the background. News, video streaming and online shopping apps are bundled with all the CN firmware. It is notable that substantially fewer third-party apps are preinstalled in the Redmi (Global) and Realme (Global) OS distributions.
\vspace{-0.5em}
\begin{tcolorbox}[standard jigsaw, arc=0pt, opacityback=0, boxrule=0.5pt,left=3pt,right=3pt]
\textit{\textbf{Key findings:} Product bundling in CN firmware is more extensive than in global firmware, and the CN firmware preinstalls multiple applications of the same type.}
\end{tcolorbox}
\vspace{-1.5em}
\begin{figure*}[t]
    \centering
    \vspace{-1em}
    \includegraphics[width=0.95\linewidth]{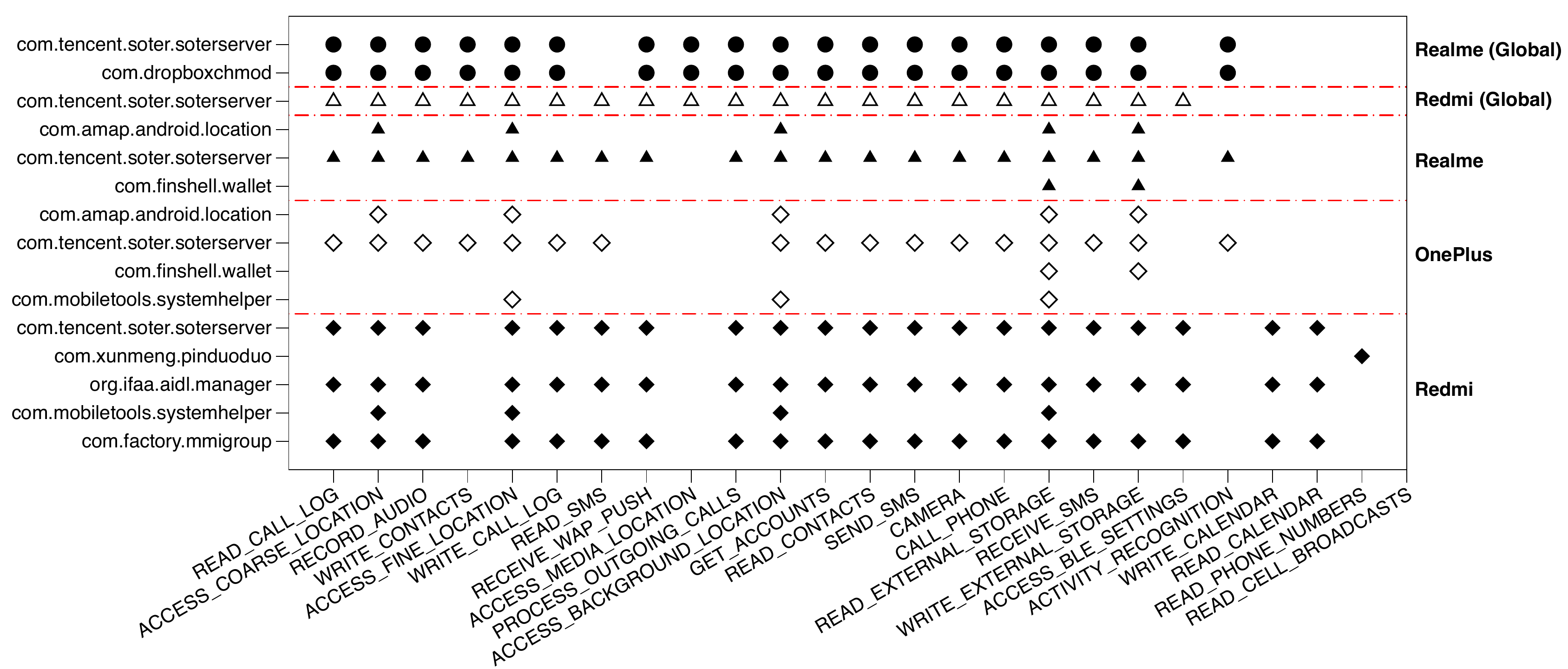}
    \vspace*{-1.5em}
    \caption{Diagram showing the runtime permissions granted by default to third-party packages by each handset.}
    \label{fig:grant_runtime}
    \vspace*{-1.5em}
\end{figure*}


\begin{tcolorbox}
Q2. How many permissions are requested by system services and preinstalled third-party apps?
\end{tcolorbox}
\vspace{-0.5em}
We can see from Table \ref{table:pkg_num} that the Android packages and vendor packages in the CN distribution request roughly the same number of permissions as those in the Global distribution.  It can also be seen that the permissions requested by vendor packages are more than twice as many as those requested by Android packages. A previous study \cite{wu2013impact} found that more than 80\% of preloaded vendor packages are over-privileged, which is consistent with our own measurements. For example, the package for fingerprint authentication \texttt{com.goodix.fingerprint} requires permission to access Calendar, Camera, Contact, Call log and Audio recording. This appears to be a common feature across different firmware. 

On the other hand, third-party packages preinstalled in all the handsets with CN distributions request many more permissions on average than Global distributions. The reason, apart from over-privileging, may be that the bundled apps request a number of permissions declared by different phone manufacturers to ensure maximal compatibility of the apps. For example, \texttt{com.taobao.taobao} on OnePlus and Realme request permissions declared by Meizu, Google, Samsung, Vivo and Huawei, although this package is installed in ColorOS. We include a detailed list of all the requested permissions and their frequency in Appendix \ref{sec:appendix}.  

Given that some custom permissions declared by different vendors serve the same purpose and could be potentially distracting, we take a closer look into the so-called dangerous ones, as defined in the official Android documentation \cite{android-doc} and which require runtime consent from users by default. In Figure \ref{fig:perm_ecdf} we plot the Empirical Cumulative Distribution Function (ECDF) of the number of requested dangerous permissions with respect to each type of preinstalled packages. A common observation in the first four subplots is a surge in the `vendor' curve at $x \sim 20$ from $\sim$ 0.75 to 1, meaning that 25\% of the vendor packages request around 20 runtime permissions. An exception is Realme (Global) which has a version of mobile OS (realme UI v1.0) that is older than the Realme (CN) (realme UI v2.0), and package over-privilege seems more prevalent. Third-party packages on CN firmware also follow a similar pattern in that around 10\% of them ask for no dangerous permissions, while the rest varies within 20 permissions. However, third-party packages on Global firmware request many fewer dangerous permissions, with 60\% on Redmi Global and 50\% on Realme Global  not using any dangerous permissions.  
\vspace{-0.5em}
\begin{tcolorbox}[standard jigsaw, arc=0pt, opacityback=0, boxrule=0.5pt,left=3pt,right=3pt]
\textit{\textbf{Key findings:} Vendor packages are over-privileged in both the CN and Global distributions. In CN distributions, pre-installed third-party apps ask for a significantly larger collection of permissions, including dangerous permissions, than in Global distributions.}
\end{tcolorbox}

\vspace{-1em}
\begin{tcolorbox}[left=7pt,right=7pt]
Q3. Do handset manufacturers grant dangerous (runtime) permissions by default to preinstalled third-party packages?
\end{tcolorbox}
\vspace{-0.5em}
The number of requested permissions, to some extent, reflects the privacy awareness of app developers, but it is not necessarily an indicator of the privacy awareness of phone manufacturers. To confirm the latter, we also study whether runtime permissions would be granted to preinstalled third-party apps by default, without user interactions, which is under the control of the vendor. Note that third-party apps which circumvent permission systems on their own are not in the scope of our analysis, while interested readers can refer to \cite{reardon201950}. We factory reset each handset to ensure that no user interaction is made with any applications, and then monitor the runtime permissions of each package via \texttt{dumpsys}. In Figure \ref{fig:grant_runtime}, we report the third-party apps that are granted runtime permissions by default, and the number of runtime permissions granted on each handset. It can be seen that \texttt{com.tencent.soter.soterserver}, an authentication package for WeChat Pay \cite{soter}, is installed on all of the handsets and is automatically granted more than 17 runtime permissions, including access to location, recording audio, reading SMS and using the camera. A payment authentication module may use such permissions to verify user identity and receive one-time tokens. However, a range of seemingly unnecessary permissions are also granted by the handsets, including reading/writing call logs and reading the list of contacts. The \texttt{org.ifaa.aidl.manager} app is an equivalent for Alipay (another popular electronic payment system in China). Moreover, on OnePlus and Realme, the \texttt{com.amap.android.location} (Amap) app is permitted to access background location without user consent, and in our analysis in Section \ref{sec:traffic_analysis}, we find that this package transmits GPS coordinates periodically to the relevant backend server.  \texttt{com.mobiletools.} \texttt{systemhelper} is a China Unicom device-registration SDK, which is also permitted to access in the background the location of the handsets. Our traffic analysis in Section \ref{sec:traffic_analysis} further reveals that a range of PII is transmitted by this package after factory reset. 

\vspace{-1em}
\begin{tcolorbox}[standard jigsaw, arc=0pt, opacityback=0, boxrule=0.5pt,left=3pt,right=3pt]
\textit{\textbf{Key findings:} Third-party packages are granted dangerous permissions by default, without the need for user interactions, resulting in user privacy exposure risks.  The user is not informed and so may be completely unaware of this.}
\end{tcolorbox}

\vspace{-1.5em}
\subsection{Code Instrumentation \& Traffic Analysis}
\label{sec:traffic_analysis}
We adopt a collection of techniques, including apk reverse engineering and dynamic runtime hooking, to aid in our analysis of the contents of the collected network traffic traces. Although most HTTP requests are on top of the TLS layer, we identify that an extra layer of content encryption/encoding is often applied when PII is uploaded to a backend server. Some HTTP requests would embed the name of the package in headers or queries, which allows us to pinpoint the source APK files and figure out the applied encryption algorithm via decompiling. However, it is also common to see HTTP requests with only a minimal set of headers and everything else encrypted. In cases where this type of requests occur, such as factory reset or opening an app, we log them and configure \textit{mitmproxy} to delay these requests for 20 seconds. We then trigger the handset to generate those requests again. Once they are observed and delayed on the proxy server side, we run exhaustive search in all the \texttt{/proc/<pid>/net/tcp6} files, which record the socket usage of each process, including source and destination IP addresses, and the UID of the connection-initiating package. After repeating the procedure on all the HTTPS requests with unreadable contents, we uncover a few data encryption routines that the app developers tend to use in practice, as summarized below. Every TCP connection can be thus attributed to an app by associating the UID with an (IP, port) pair. We note that while it is theoretically feasible to transmit covert messages in DNS requests, in practice such requests may never reach the DNS servers hosted by app owners, due to the hierarchical DNS resolution architecture in the network.

\vspace{-1.5em}
\paragraph{Symmetric Encryption}
We identify a large number of packages that apply AES encryption algorithms when uploading PII, by decompiling the associated APK files. For example, \texttt{com.coloros.} \texttt{weather} on Realme and OnePlus encrypts GPS coordinates and PII by AES (CTR mode), and concatenates the key and the initialization vector at the end of the ciphertext. \texttt{com.nearme.romupdate} also appends AES keys at the end when querying system update information. On the other hand, \texttt{com.ted.number}, China Unicom SDK, \texttt{com.nearme.instant.platform} and \texttt{com.android.} \texttt{updater} (Xiaomi) hardcode AES keys in the package, which are used to encrypt POST contents of a range of HTTP requests. For these packages using symmetric encryption algorithms, as long as we acquire the AES keys, the contents can be easily decrypted. 

However, an exception is \texttt{com.amap.android.location} which encapsulates encryption/decryption algorithm in a precompiled JNI library. Instead of getting the keys, we have to hook the plaintext before encryption during runtime, for which we utilize Frida and Riru/EdXposed. Frida \cite{frida} is a dynamic code instrumentation tool that has full access to the memory and injects Javascript code into the target process on Android, allowing users to hook variables and function calls during runtime. 
However, it is not possible to spawn an Android process that has no foreground activity, meaning that hooking the variables created right after a process is started can be difficult. An alternative is using Riru/EdXposed. Riru \cite{riru} injects code into Zygote which is a special process handling the forking of every new Android app, and EdXposed \cite{edxposed} is an Riru module that provides universal hooking APIs. This framework is able to inject code into the app at the very beginning when it is forked from Zygote, thereby providing better control of code instrumentation than Frida. 
In this work we use Frida 15.2.2, Riru 25.4.4 and EdXposed 0.5.2.2.

\paragraph{Asymmetric Encryption}
Asymmetric encryption is involved in a small number of packages, such as \texttt{com.mobiletools.system} \texttt{helper} (China Unicom SDK), when uploading private information. Specifically, the package contains a hardcoded RSA public key, and each HTTP request contains the ciphertext encrypted with an AES key and also the RSA-encrypted AES key. Since the RSA private key is stored on the backend server, we have to utilize Frida to hook runtime plaintext before it gets encrypted. A few Xiaomi packages adopt similar routines while uploading app telemetry.

\begin{table}[t]
\bgroup
\def\arraystretch{1.1}
\begin{tabular}{llccccc}
\Xhline{2\arrayrulewidth}
\multicolumn{2}{c}{PII type} & 
\multicolumn{1}{P{90}{1cm}}{Redmi}  & 
\multicolumn{1}{P{90}{1cm}}{OnePlus}  & 
\multicolumn{1}{P{90}{1cm}}{Realme}  & 
\multicolumn{1}{P{90}{1cm}}{Redmi (Global)} & 
\multicolumn{1}{P{90}{1cm}}{Realme (Global)}\\ \Xhline{2\arrayrulewidth}
\multirow{5}{*}{\begin{tabular}[c]{@{}l@{}}Device\\ specific\end{tabular}} & IMEI & \checkmark & \checkmark & \checkmark & \checkmark & \checkmark \\
 & temporary IDs & \checkmark & \checkmark & \checkmark & \checkmark & \checkmark \\
 & installed apps & \checkmark & \checkmark & \checkmark & \checkmark & \checkmark \\
 & OS version & \checkmark & \checkmark & \checkmark & \checkmark & \checkmark \\
 & hardware info & \checkmark & \checkmark & \checkmark & \checkmark & \checkmark \\ \hline
\multirow{7}{*}{\begin{tabular}[c]{@{}l@{}}Geo-\\ locaiton\end{tabular}} & MCC & \checkmark & \checkmark & \checkmark &  & \checkmark \\
 & MNC & \checkmark & \checkmark & \checkmark &  & \checkmark \\
 & CID & \checkmark & \checkmark & \checkmark &  &  \\
 & LAC & \checkmark & \checkmark & \checkmark &  &  \\
 & GPS & \checkmark & \checkmark & \checkmark &  &  \\
 & connected Wi-Fi & \checkmark & \checkmark & \checkmark & \checkmark &  \\
 & nearby Wi-Fi & \checkmark & \checkmark & \checkmark &  &  \\
 \hline
\multirow{4}{*}{\begin{tabular}[c]{@{}l@{}}User \\ profile\end{tabular}} & phone number & \checkmark & \checkmark & \checkmark &  &  \\
& IMSI & \checkmark & \checkmark &  & \checkmark &  \\
& ICCID & \checkmark & \checkmark &  &  & \\
 & app usage & \checkmark & \checkmark & \checkmark & \checkmark &  \\ \hline
\multirow{2}{*}{\begin{tabular}[c]{@{}l@{}}Social \\ relationship\end{tabular}} & SMS history &  & \checkmark & \checkmark &  &  \\
 & call history &  & \checkmark & \checkmark &  &  \\ \Xhline{2\arrayrulewidth}
\end{tabular}
\egroup
\caption{Specific types of PII uploaded by each handset. `Temporary ID' represents IDs created by vendor packages or OS, which would change after factory reset.}
\label{table:pii_upload}
\vspace*{-3.5em}
\end{table}

\begin{tcolorbox}
Q4. Which type of personally identifiable information (PII) is uploaded to backend servers?
\end{tcolorbox}
We classify important PII into four categories and in Table \ref{table:pii_upload} present which specific identifiers or information is actually shared. 

\vspace{-0.5em}
\paragraph{Device-specific PII}
It appears that the handsets studied routinely upload device-specific PII, regardless of brand and firmware. Installed applications and the version of Android OS is posted periodically to check for updates. Hardware information, including phone model, CPU brand and screen size are also posted to the backend servers. Besides, persistent identifiers, such as IMEI, and resettable identifiers, including VAID, OAID and Android ID, are embedded in a number of requests for device registration, telemetry, checking for updates, etc. However, we notice that the mobile devices with the global version of firmware only transmit the IMEI to the vendor, while those with CN version also upload the IMEI to China Unicom (\url{dm.wo.com.cn}) and China Mobile (\url{a.fxltsbl.com}) for registration after factory reset and when a SIM card is inserted, despite the handsets not having a contract with any of these communication service providers. A quick search reveals that the URLs correspond to the device management platforms for the mobile operators, and apparently any Android device in China Mobile's product library (including non-contract phones) should support the SDK for device management. Although we did not find the official regulation file, the SDK is installed on all the devices with CN firmware, named \texttt{com.miui.dmregservice} on Xiaomi and \texttt{com.coloros.regservice} on OnePlus and Realme. Different from China Unicom, the China Mobile SDK also collects and uploads the installed package list in a request to \url{https://a.fxltsbl.com/accept/sdkService}, for unknown purpose. 

Despite the same start-up configuration on Redmi and Redmi (Global) described in Section \ref{sec:dataset}, we discover in the payload to \url{api.ad.xiaomi.com} that the handsets have different system advertising configuration:

\begin{lstlisting}[escapechar=!]
!\http{POST}! https://api.ad.xiaomi.com/brand/pushConfig
clientInfo={...{"locale":"zh_CN","language":"zh","country":"CN","isPersonalizedAdEnabled":!\textbf{true}!,...}}

!\http{POST}! https://api.ad.intl.xiaomi.com/brand/pushConfig
clientInfo={...,{"locale":"en_US","language":"en","country":"IE""isPersonalizedAdEnabled":!\textbf{false}!..}}
\end{lstlisting}
That said, this feature cannot be controlled by opting out of `Send Usage Data' or `User Experience Program', but this is an inherent difference between the CN and Gloabl firmware. It is still unclear how this option would impact advertising behaviors though, which is a decision to be taken at the backend side.  

\paragraph{Geolocation PII} Device-related PII are posted to backend servers regardless of firmware, but we find that the uploading of geolocation-related PII appears substantially different between the global and CN fimware. For Realme (Global), only the MCC and MNC are uploaded to \url{confe.dc.oppomobile.com}, which allows the server to confirm the country and the mobile operator in use. Redmi (Global) sends the connected Wi-Fi SSID and MAC address to \url{tracking.intl.miui.com}, which leaks user coarse location if the server maintains a database with Wi-Fi access point information.\footnote{3WiFi: open database of Wi-Fi Access Points passwords \url{https://miloserdov.org/?p=746}}

However, for handsets with CN firmware, traffic analysis reveals that the full range of the geolocation-related PII is uploaded, and to a range of recipients. The MCC, MNC, CID and LAC are transmitted to China Mobile and Unicom when a SIM card is inserted. By using the CID and LAC, these Chinese mobile operators can therefore easily infer coarse user location. A notable fact in Table \ref{table:pii_upload} is that the three handsets with CN firmware also upload GPS coordinates.  This happens even for the Redmi handset for which we chose to disable location during device startup following factory reset (the OnePlus and Realme handsets do not offer this option and have location service turned once the devices are started). 

Both the OnePlus and Realme handsets have the \texttt{com.coloros. weather.service} and \texttt{com.amap.android} \texttt{d.location} apps preinstalled, which are granted the \textit{ACCESS\_BACKGROUND\_L -OCATION} permission by default, and upload the current GPS location periodically. 

The weather app makes the following request:
\begin{lstlisting}[escapechar=!]
!\http{POST}! https://i6.weather.oppomobile.com/weather/location/v0/sdk?appId=app-weather&authCode=3357..
{ '!\textbf{latitude}!': <anonymized>,'language': 'ZH-CN',  '!\textbf{longitude}!': <anonymized>,'vaid': '63..5B', 'oaid':..}
\end{lstlisting}
in which latitude and longitude are populated with encrypted GPS location and the associated AES keys. The request contains identifiers (OAID and VAID), which can be combined with data from other connections to identify the individual handset, and is sent to an Oppo server approximately 5 times per day. AMap encapsulates the encryption algorithm in a JNI library, and sends the ciphertext in the following message:
\begin{lstlisting}[escapechar=!]
!\http{POST}! http://apsrgeo.amap.com/rgeo/r?q=1&language=cn
{"data":"...<!\textbf{latitude}!><anonymized><\/latitude><!\textbf{longitude}!><anonymized><\/longitude>..."}
\end{lstlisting}
The connections to this domain are recorded 13 times a day, with an ID \textit{adiu} associated, allowing Amap's server to analyze the user's location over time. Moreover, AMap also transmits the LAC, CID, the connected Wi-Fi MAC and SSID, as well as nearby Wi-Fi MACs and SSIDs in the request shown below:
\begin{lstlisting}[escapechar=!]
!\http{POST}! http://aps.amap.com/APS/r?ver=5.1&q=0&csid=d6...
{"!\textbf{mcc}!":<anonymized>,"!\textbf{mnc}!":<anonymized>,"!\textbf{lac}!":<anonymized>,"!\textbf{cid}!":<anonymized>,"wifiStatus":{...,"!\textbf{mainWifi}!":{"mac":"<anonymized>","ssid":"\"androidprivacy\"
", "!\textbf{wifiList}!":[{"mac":"<anonymized>","ssid
":"Xiaomi_8DEE",...}...]}
\end{lstlisting}
where \textit{androidprivacy} is the Wi-Fi network we deployed for traffic collection purposes and \textit{Xiaomi\_8DEE} is a nearby WiFi. Note that the actual payload is encrypted and posted in a compact format, in which only the values are populated in a byte array. 

In addition, the \texttt{com.coloros.wifisecuredetect} app on the OnePlus and Realme handset transmits the connected Wi-Fi SSID and MAC address to \url{log.avlyun.com}.
\begin{lstlisting}[escapechar=!]
!\http{POST}! https://log.avlyun.com/logupload?channel=coloros_wifi&pkg=com.coloros.wifisecuredetect
#SYSINFO;{"kernel_version":"","name":"OnePlus9R_CH","security_patch":"2021-03-05","sdk":"30","incremental":"1618459936301","base_os":"","platform":"kona","manufacturer":"OnePlus"}
#WIFI;635eb9bf(timestamp in hex); !\textbf{androidprivacy}! (wifi name);<!\textbf{anonymized}!> (Mac address);PSK;;;...
\end{lstlisting}

The Redmi handset does not grant runtime permissions to the map and weather apps by default, but allows system services to access location information in the background after startup. The \texttt{com.miui.analytics} system app sends the MCC, MNC, GPS coordinates and nearby Wi-Fi access point to \url{https://tracking.miui.com/track/v4} as shown below:
\begin{lstlisting}[escapechar=!]
!\http{POST}! https://tracking.miui.com/track/v4
{"!\textbf{radio}!": "<anonymized>",  // MCC+MNC
"loc": "{\"w\":\"<!\textbf{anonymized\_MAC\_addr}!>,!\textbf{androidprivacy}!,-37,5500\",...\"wl\":[\"<!\textbf{anonymized\_MAC\_addr}!>,!\textbf{Xiaomi\_8DEE}!,-31,2422\",...,\"!\textbf{lat}!\":<anonymized>,\"!\textbf{lng}!\":<anonymized>}
\end{lstlisting}
and \texttt{com.xiaomi.metoknlp} encrypts and transmits the GPS coordinates to Baidu Map API to retrieve street information:
\vspace{-0.25em}
\begin{lstlisting}[escapechar=!]
!\http{GET}! https://api.map.baidu.com/geocoder/v2/?channel=nl.1269e&coordtype=wgs84ll&&from=BaiduNLP&!\textbf{location}!=<anonymized>
\end{lstlisting}
Both services run in the background.  The former sends geolocation PII right after a SIM card is inserted, while the latter is observed once within the first 24 hours of leaving the device idle after factory reset. Note however that we found no device/user identifiers in the request to Baidu.


\paragraph{User profile PII}
We group user phone number and ICCID (SIM card identifier) as a part of user profile information because every phone number in China is registered under a citizen ID, thus functions as a semi-persistent identifier of the user. Due to the existence of China Unicom SDK on Redmi and OnePlus, phone number, IMSI and ICCID are transmitted within the first 10 minutes after factory reset. The Phone and Message apps on OnePlus and Realme would send contact information when receiving text/calls, which is intended for ``recognizing unknown number'', but also leaks app usage information because of the timestamp and call duration in such requests, as detailed in what follows. 

On the other hand, Redmi collects app usage data from a much broader range with greater level of detail. We uncover that Redmi posts requests to \url{tracking.miui.com/track/v4} when the preinstalled Settings, Note, Recorder, Phone, Message and Camera apps are opened and used. The app's first launch time, usage start time, end time, and the timestamps when accessing some Android activities, such as WifiProvisionSettingsActivity, NotesList Activity, EditActivity and Camera, are uploaded. If the user is accessing preinstalled Xiaomi apps consecutively, telemetry would be logged frequently, resulting in a chain of actions that the user conducts. A snippet of the relevant request is shown below:
\begin{lstlisting}[escapechar=!]
!\http{POST}! https://tracking.miui.com/track/v4
{ "imsis": "[b2d5c6783e3fa6eef38ff1fc7dedfb10,]",..,
{"pkg": "com.xiaomi.smarthome","action": "first_launch", "fit": 1666816796000, ...},
{"pkg": "com.android.settings","ts": 1666818456958,"duration": 1424, ...},
{"pkg": "com.miui.securityinputmethod","ts": 1666818463544,"duration": 4706, ... },
{"pkg": "com.miui.notes","ts": 1666818784908,"stat": "app_start",...}...}
\end{lstlisting}
This type of telemetry is not affected by opting out of `Send Usage and Diagnositic Data' during start-up. 

It was previously reported that Redmi (Global) also collects IMSI and app telemetry, but the other user profile PII seen above are not transmitted from Redmi (Global) and Realme (Global) \cite{our-report}.

\paragraph{Social relationships PII} 
The preinstalled Phone (\texttt{com.ted.numb} \texttt{er}) and Message (\texttt{com.android.mms}) apps on OnePlus and Realme not only transmit user's phone number, but also send the other party's phone number with the duration when making/receiving a phone call or sending/receiving a text message. A typical request generated with a phone call is shown below:
\begin{lstlisting}[escapechar=!]
!\http{POST}! https://sms.ads.heytapmobi.com/new/v5/phone
{"header":{.."p7":"!\textbf{user-phone-number}!",...,"data":{"phone":"!\textbf{caller-number}!","dialNumber":"!\textbf{caller-number}!",...,"!\textbf{duration}!":15,"contact":-1,"ringtime":0,"lasttime":-1}}  
\end{lstlisting}
in which call duration, ring time, last contact time and whether the caller is in the contact is posted to the backend server. This type of PII not only helps the vendors to identify individual users, but also leaks users' social relationships. For the group of population who are inclined towards these brands, manufacturers can even build a connection map among them on the backend side, and infer the social relationships between users who are not directly connected. 
\begin{tcolorbox}[standard jigsaw, arc=0pt, opacityback=0, boxrule=0.5pt,left=3pt,right=3pt]
\textit{\textbf{Key findings:} CN OS distributions transmits a much larger range of PII to backend servers than Global distribution do, despite the fact that they are developed by the same companies. This is facilitated by 1) the granting of dangerous permissions to some pre-installed apps by default; and 2) a number of preloaded third-party apps being allowed to run continuously in the background.}
\end{tcolorbox}

\begin{tcolorbox}
Q5. Is the user notified about the transmission of PII regarding geolocation, user profile and social relationships?
\end{tcolorbox}
During device start-up following factory reset, each handset may present the user with information on the permissions used and data collected by preinstalled apps.  Also, the first time an app is opened the terms and conditions of the specific app may be shown, and permissions may be asked. Having observed the data actually sent from each handset, we now revisit the information provided to users to check whether this matches the actual data transmission we observe. 

\vspace{-0.5em}
\paragraph{Xiaomi} The following system apps transmit geolocation and user profile PII:

    1) \texttt{com.miui.anlaytics}: The statement in the startup page shows that Analytics would use location service and transmits the IMSI. ``If you don't agree to grant such permissions ... You can choose not to use this feature in such cases.'' However, no option is given to restrict the permissions of this app and this app cannot be managed in Settings > Apps.

    2) \texttt{com.xiaomi.metoknlp}: We did not find any relevant statement for this app, which is granted \textit{ACCESS\_BACKGROUND\_L-} \textit{OCATION} and posts GPS coordinates to the Baidu Map API.
    
    3) \texttt{com.mobiletools.systemhelper}: This app embeds the China Unicom SDK. A statement on carrier services gives the types of information (including coordinates) to be collected.
    
    4) \texttt{com.miui.regservice}: This embeds the China Mobile SDK. Similarly to the above, a statement on carrier services details the types of information (including coordinates) that will be collected by China Mobile.

A shared feature of these statements is that the user is presented with a `take-it-or-leave-it' choice by the vendor and mobile operators together, and the user is unable to fully control the permissions of preinstalled packages. For example, if the user would like to avoid GPS coordinates being uploaded, the only option is to turn Location off in `Settings', which applies to all apps and so likely to cause significant inconvenience. Note also that when starting the Redmi handset we chose to disable Location, but we found that the location service is still running once we enter the system. 

\vspace{-0.5em}
\paragraph{OnePlus \& Realme}
Since both handsets are shipped with Color\-OS, PII transmission on these devices share a large degree of similarity. The following apps transmit geolocation, user profile and social relationships-related PII:

    1) \texttt{com.coloros.weather.service}: The statement on the startup page shows that ``during your use of the `Weather Service' [...], we need to use your location information''. However, the Weather app is never opened during our data collection campaign. 
    
    2) \texttt{com.amap.android.location}: We did not find any relevant statement for this app to access location information.
    
    3) \texttt{com.ted.number}: When opening the Phone app for the first time, the user is asked to agree to terms, including permissions to read call logs to identify unknown numbers. However, the user is not offered the option to decline these terms. 
    
    4) \texttt{com.android.mms}: The above situation also applies to the Message app. 
    
    5) \texttt{com.coloros.regservice}: This embeds the China Mobile SDK. The startup statement informs users that this package has the permission to access location information and would regularly upload PII to the backend server.
    
    6) \texttt{com.mobiletools.systemhelper} (OnePlus only): This package embeds the China Unicom SDK. A statement about the carrier's term of service details the information to be collected.

Compared with Xiaomi, we find that OnePlus and Realme are even less transparent about the use of location information. It should be noted that the Weather Service and Weather are two different packages, in which the former has \textit{ACCESS\_BACKGROUND\_LO}\textit{-CATION} granted by default and runs in the background, and thereby we can observe coordinates being uploaded. However, if the user checks the permission status of Weather in Settings > Permission Manager, access to location information is never granted because it is the Weather Service that has the runtime permission. Other packages also trap users in a ``take-it-or-leave-it'' situation, where a `decline' button is not provided. 
\vspace{-0.5em}
\begin{tcolorbox}[standard jigsaw, arc=0pt, opacityback=0, boxrule=0.5pt,left=3pt,right=3pt]
\textit{\textbf{Key finding:} In general, the user is not notified about the transmission of important PII or is not given the option to reject such transmissions when being notified.}
\end{tcolorbox}

\begin{figure}[t]
    \vspace{-1em}
    \centering
    \includegraphics[width=\columnwidth]{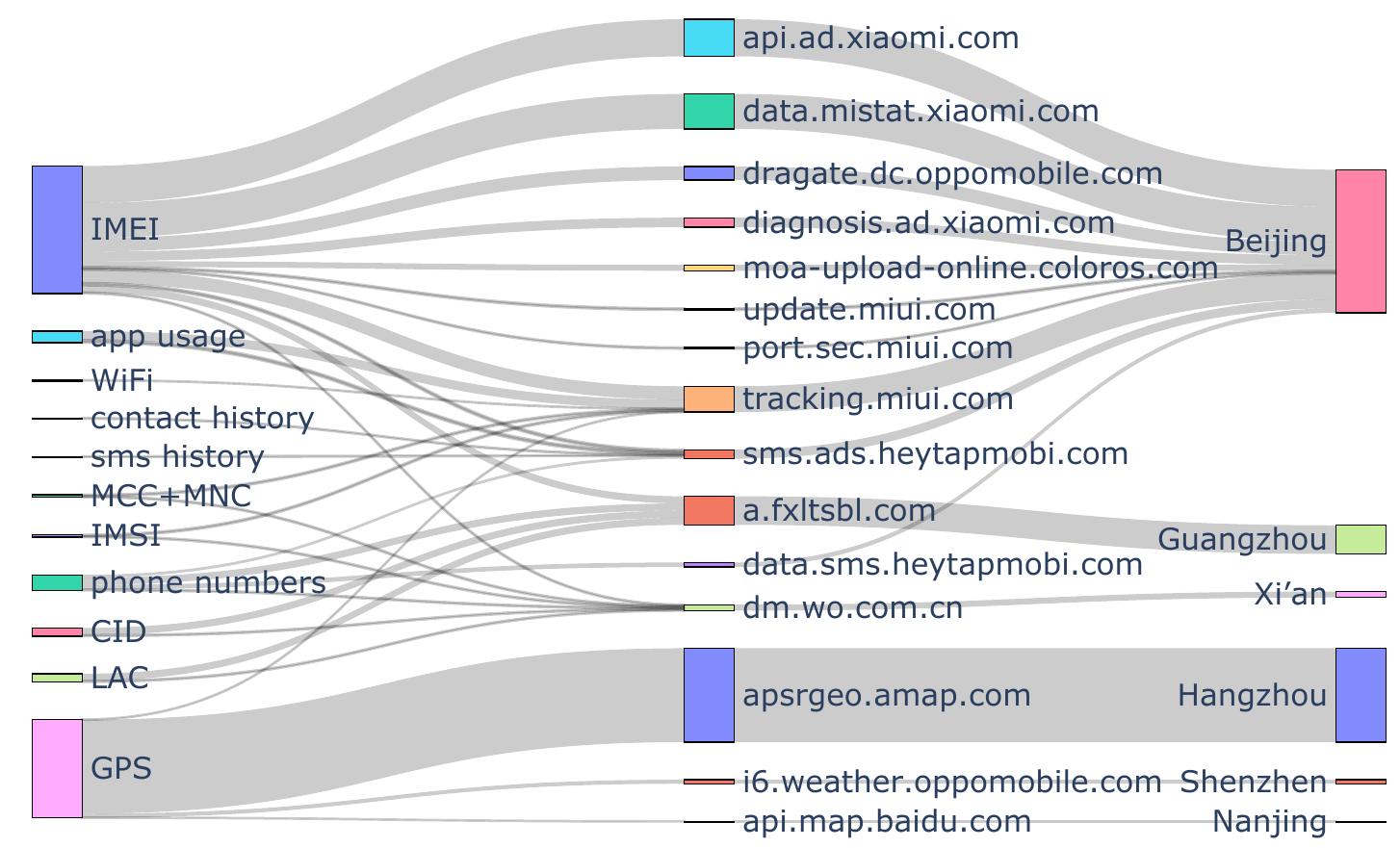}
    \vspace{-1.5em}
    \caption{Sankey Diagram of important PII collected by the handsets running CN firmware and by what package/to what location it is being sent.}
    \label{fig:cn_sankey}
    \vspace{-2em}
\end{figure}

\begin{figure}[t]
    \centering
    \vspace{-1em}
    \includegraphics[width=\columnwidth]{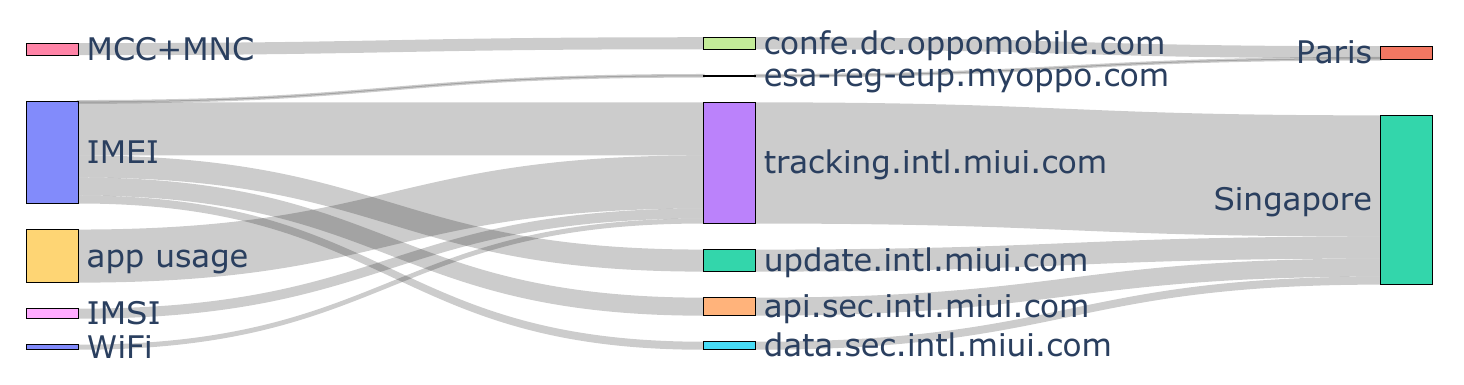}
    \vspace{-2.5em}
    \caption{The Sankey Diagram of important PII collected by the handsets with Global firmware.}
    \label{fig:global_sankey}
    \vspace{-1.5em}
    \end{figure}

\vspace{-1em}
\begin{tcolorbox}
Q6. Who collects PII and where are they located?
\end{tcolorbox}
Figure \ref{fig:cn_sankey} and Figure \ref{fig:global_sankey} visualize, respectively, the PII collection in the CN and Global distributions. For the same device setup after factory reset, it can be clearly seen that the handsets running CN firmware collect a wider range of PII and transmit to more endpoints owned not only by phone vendors but also by third-party domains, in which the IMEI is the most frequently collected persistent identifier. The AMap system app on the Realme and Oneplus handsets regularly transmits GPS coordinates when the devices are idle, ranking the second in Figure \ref{fig:cn_sankey}. On handsets running CN firmware, the phone vendors, weather and navigation providers and mobile operators all collect important PII in the background. It is not surprising to see that all of the servers associated with the endpoints are located in China, since the China Data Protection Law explicitly restricts personal data from being transferred abroad without administrative consent. For handsets running the Global firmware, the IMEI is also the most frequently transmitted persistent identifier, followed by telemetry data collected by the Redmi Global firmware. However, we do not find that any preinstalled third-party apps collecting geolocation, user profiles or social-relationship PII. It should be noted that the measurement traces for the Redmi Global firmware were collected one year ago when DNS resolution of Xiaomi's domain pointed to IP addresses in Singapore, which may be different from newer DNS records. 

\vspace{-0.5em}
\begin{tcolorbox}[standard jigsaw, arc=0pt, opacityback=0, boxrule=0.5pt,left=3pt,right=3pt]
\textit{\textbf{Key finding:} Important PII transmitted by CN firmware reaches many third-party domains, whereas this is not witnessed with Global firmware.}
\end{tcolorbox}

\vspace{-1em}
\section{Related Work}
\paragraph{Android Customization Analysis}
Previous work on Android customization analysis mainly focuses on the security side \cite{aafer2016harvesting}. It was shown that vulnerable security configurations exist at a large scale in custom Android firmware, and these may potentially lead to a series of privacy attacks, such as stealing emails and altering system settings without proper permissions \cite{aafer2016harvesting}. Wu et al. found that due to vendor customization, more than 80\% of pre-loaded packages are over-privileged, some of which can be exploited for permission re-delegation attack or privacy leaks \cite{wu2013impact}. The customization of device drivers by Samsung is found to underpin various attacks, including taking a photo or screenshot without permissions \cite{zhou2014peril}.

\paragraph{Privacy leak analysis through app monitoring:}
Mobile apps are capable of accessing a series of PII with or without user permission. It can be extremely time-consuming to manually examine which type of PII is accessed by each app, thus automated tools come into play. PiOS analyzes control flow graphs of binary files and examines the reachability of key PII for iOS applications \cite{egele2011pios}. FlowDroid proposes a static analysis method by taking the Android Application life-cycle into consideration \cite{arzt2014flowdroid}. AppIntent studies if the transmission of private data is intended by the users through symbolic execution \cite{yang2013appintent}. Taintdroid \cite{enck2014taintdroid} marks (taints) sensitive data and tracks the data flow during execution. However, these tools face challenges when PII is obfuscated or encrypted. Continella et al. \cite{continella2017obfuscation} proposes a black-box method without the knowledge of app source code to detect privacy leaks through differential analysis. 

\vspace{-0.5em}
\paragraph{Traffic Analysis:}
It has been shown that a vast amount of PII is directly embedded in queries, headers or post body of HTTP requests. By using ISP traffic logs or self-collected datasets, numerous mobile advertising and tracking services were uncovered, in which persistent identifiers including IMEI, IMSI, MAC address and advertising IDs are often transmitted \cite{vallina2012breaking, wang2020exploring, narseo18, nguyen2021share}. Note that most of the services are hosted by third-party companies instead of phone vendors. Ren et al. track the update of more than 500 apps across 8 years and document the evolution of PII collection \cite{ren2018bug}. Further work examines over 500 apps in the Google Play Store and shows that 76\% of them collect and transmits PII insecurely and 34\% send PII to third parties \cite{jia2019leaks}. Privacy analyses have also been conducted to compare over 5,000 free mobile apps with their paid versions, demonstrating that paid apps are not necessarily more privacy-aware than their free counterparts, with 34\% exhibiting the same data transmission behavior \cite{han2020price}. 
Moreover, a case study on the Covid contact tracing apps was conducted recently, revealing that the Google Play Services integrated in the apps would regularly contact Google servers, making it possible to track users' location via the change of IP addresses \cite{contacttrace}. It was reported that both Google and Apple collect a number of PII including IMEI, IMSI and telemetry. iOS devices also transmit nearby Wi-Fi MAC addresses and GPS coordinates, which users cannot control \cite{iosgoogle}. Previous work analyzes the data collection of different Android devices (Huawei, Xiaomi and Oppo) with global firmware \cite{our-report}, while our study fills the gap of firmware differences across regions in terms of PII collection.

\vspace{-0.5em}
\section{Conclusion}
In this work we study the Chinese version of the Android OS distributions run by Xiaomi, Realme, and OnePlus handsets.  We measure the network traffic the handsets generate when in-use by a privacy-aware consumer. We find that these devices come bundled with a number of third-party applications, some of which are granted dangerous runtime permissions by default without user consent, and transmit traffic containing a broad range of geolocation, user-profile and social relationships PII to both phone vendors and third-party domains, without notifying the user or offering the choice to opt-out. In contrast, the data shared by the Global version of the firmware is mostly limited to device-specific information.  Our study therefore highlights major differences in terms of how privacy provisions are enforced in different regions.

\section*{Acknowledgments}
This material is based upon work partially supported
by Arm Ltd and Scotland's Innovation Centre for sensing, imaging and Internet of Things technologies (CENSIS).

\bibliographystyle{ACM-Reference-Format}
\balance
\bibliography{bibfile}

\begin{appendices}

\section{Appendix}
\label{sec:appendix}
The list of permissions requested by third-party packages are summarized in Table \ref{table:freq_perms_CN} and \ref{table:freq_perms_global}. The number at the end of each entry denotes the average times that this permission is requested by third-party packages per handset. Table \ref{table:freq_perms_CN} only shows a subset of permissions which occur over 7 times on average. 
\begin{table*}[t]
\bgroup
\def\arraystretch{1.2}
\begin{tabular}{ll}
\Xhline{2\arrayrulewidth}
permission & permission \\ \hline
android.permission.INTERNET (26.7) & android.permission.ACCESS\_NETWORK\_STATE (26.7)\\
android.permission.READ\_EXTERNAL\_STORAGE (25.7) & android.permission.ACCESS\_WIFI\_STATE (25.3)\\
android.permission.WRITE\_EXTERNAL\_STORAGE (25.0) & android.permission.READ\_PHONE\_STATE (24.7)\\
android.permission.VIBRATE (23.0) & android.permission.WAKE\_LOCK (22.7)\\
android.permission.REQUEST\_INSTALL\_PACKAGES (22.0) & android.permission.ACCESS\_FINE\_LOCATION (21.7)\\
android.permission.ACCESS\_COARSE\_LOCATION (21.7) & android.permission.CAMERA (21.7)\\
android.permission.GET\_TASKS (21.3) & android.permission.CHANGE\_NETWORK\_STATE (21.0)\\
android.permission.RECORD\_AUDIO (21.0) & android.permission.CHANGE\_WIFI\_STATE (20.3)\\
com.android.launcher.permission.INSTALL\_SHORTCUT (19.3) & android.permission.FOREGROUND\_SERVICE (19.3)\\
android.permission.SYSTEM\_ALERT\_WINDOW (19.3) & android.permission.WRITE\_SETTINGS (18.0)\\
android.permission.BLUETOOTH (17.7) & com.coloros.mcs.permission.RECIEVE\_MCS\_MESSAGE (17.0)\\
android.permission.MODIFY\_AUDIO\_SETTINGS (16.7) & com.android.launcher.permission.READ\_SETTINGS (15.7)\\
android.permission.RECEIVE\_BOOT\_COMPLETED (15.0) & android.permission.READ\_CONTACTS (15.0)\\
com.android.launcher.permission.UNINSTALL\_SHORTCUT (14.3) & android.permission.FLASHLIGHT (13.7)\\
android.permission.READ\_LOGS (13.3) & com.huawei.android.launcher.permission.CHANGE\_BADGE (13.0)\\
android.permission.MOUNT\_UNMOUNT\_FILESYSTEMS (12.7) & com.oppo.launcher.permission.READ\_SETTINGS (12.3)\\
android.permission.USE\_FINGERPRINT (12.0) & android.permission.REORDER\_TASKS (11.7)\\
com.htc.launcher.permission.READ\_SETTINGS (11.3) & com.huawei.android.launcher.permission.READ\_SETTINGS (11.0)\\
android.permission.USE\_CREDENTIALS (11.0) & com.oppo.launcher.permission.WRITE\_SETTINGS (10.7)\\
android.permission.BLUETOOTH\_ADMIN (10.7) & android.permission.EXPAND\_STATUS\_BAR (10.3)\\
com.huawei.android.launcher.permission.WRITE\_SETTINGS (10.0) & android.permission.WRITE\_CALENDAR (10.0)\\
android.permission.GET\_ACCOUNTS (10.0) & com.meizu.flyme.push.permission.RECEIVE (9.7)\\
com.meizu.c2dm.permission.RECEIVE (9.7) & android.permission.MANAGE\_ACCOUNTS (9.7)\\
com.heytap.mcs.permission.RECIEVE\_MCS\_MESSAGE (9.3) & android.permission.CHANGE\_WIFI\_MULTICAST\_STATE (9.3)\\
com.sec.android.provider.badge.permission.WRITE (9.3) & com.sec.android.provider.badge.permission.READ (9.3)\\
com.vivo.notification.permission.BADGE\_ICON (9.3) & android.permission.AUTHENTICATE\_ACCOUNTS (9.3)\\
android.permission.READ\_CALENDAR (9.0) & com.sonyericsson.home.permission.BROADCAST\_BADGE (9.0)\\
com.android.launcher.permission.WRITE\_SETTINGS (8.7) & android.permission.WRITE\_SYNC\_SETTINGS (8.0)\\
android.permission.BROADCAST\_STICKY (8.0) & com.android.launcher3.permission.READ\_SETTINGS (8.0)\\
com.bbk.launcher2.permission.READ\_SETTINGS (8.0) & com.htc.launcher.permission.UPDATE\_SHORTCUT (7.7)\\
android.permission.DISABLE\_KEYGUARD (7.0) & com.sonymobile.home.permission.PROVIDER\_INSERT\_BADGE (7.0)\\
android.permission.ACCESS\_LOCATION\_EXTRA\_COMMANDS (7.0) & \\\Xhline{2\arrayrulewidth}
\end{tabular}
\egroup
\caption{The most frequently requested permissions by preinstalled third-party packages in the CN firmware. }
\label{table:freq_perms_CN}
\end{table*}

\begin{table*}[b]
\vspace{-1em}
\bgroup
\def\arraystretch{1.2}
\begin{tabular}{ll}
\Xhline{2\arrayrulewidth}
permission & permission \\ \hline
android.permission.ACCESS\_NETWORK\_STATE (3.3) & android.permission.INTERNET (3.3)\\
android.permission.WAKE\_LOCK (2.7) & android.permission.RECEIVE\_BOOT\_COMPLETED (2.7)\\
android.permission.READ\_EXTERNAL\_STORAGE (2.3) & android.permission.ACCESS\_WIFI\_STATE (2.3)\\
android.permission.WRITE\_EXTERNAL\_STORAGE (2.0) & android.permission.READ\_PHONE\_STATE (1.3)\\
com.google.android.c2dm.permission.RECEIVE (1.3) & android.permission.FOREGROUND\_SERVICE (1.3)\\
android.permission.VIBRATE (1.0) & com.google.android.finsky.permission.BIND\_GET\_INSTALL\_REFERRER\_SERVICE (1.0)\\
android.permission.REQUEST\_INSTALL\_PACKAGES (1.0) & android.permission.ACCESS\_MEDIA\_LOCATION (1.0)\\
android.permission.GET\_TASKS (1.0) & android.permission.INSTALL\_PACKAGES (1.0)\\\Xhline{2\arrayrulewidth}
\end{tabular}
\egroup
\caption{The requested permissions by preinstalled third-party packages in the Global firmware.}
\label{table:freq_perms_global}
\end{table*}

\onecolumn

\end{appendices}

\end{document}